\documentclass[aps,reprint,superscriptaddress,prl]{revtex4-2}
\usepackage[utf8]{inputenc}
\usepackage[T1]{fontenc}
\usepackage{amsmath}
\usepackage{amssymb}
\usepackage{graphicx}
\graphicspath{{figures/}}
\usepackage{color}
\usepackage[bookmarks=false]{hyperref}
\usepackage[usenames,dvipsnames]{xcolor}
\usepackage{subfigure}
\usepackage[super]{nth}
\usepackage[normalem]{ulem}

\def\be{\begin{equation}}
\def\ee{\end{equation}}
\def\ba{\begin{eqnarray}}
\def\ea{\end{eqnarray}}

\begin{document}


\title{Interplay between magnetism and superconductivity in UTe$_2$ }


\author{Di S. Wei}
\affiliation{Geballe Laboratory for Advanced Materials, Stanford University, Stanford, CA 94305}
\affiliation{Department of Applied Physics, Stanford University, Stanford, CA 94305}
\affiliation{Stanford Institute for Materials and Energy Sciences, SLAC National Accelerator Laboratory, 2575 Sand Hill Road, Menlo Park, CA 94025}

\author{David Saykin}
\affiliation{Geballe Laboratory for Advanced Materials, Stanford University, Stanford, CA 94305}
\affiliation{Department of Physics, Stanford University, Stanford, CA 94305}
\affiliation{Stanford Institute for Materials and Energy Sciences, SLAC National Accelerator Laboratory, 2575 Sand Hill Road, Menlo Park, CA 94025}

\author{Oliver Y. Miller}
\affiliation{Geballe Laboratory for Advanced Materials, Stanford University, Stanford, CA 94305}
\affiliation{Department of Physics, Stanford University, Stanford, CA 94305}

\author{ Sheng Ran }
\affiliation{Department of Physics, Quantum Materials Center, University of Maryland, College Park, MD 20742, USA.}
\affiliation{NIST Center for Neutron Research, National Institute of Standards and Technology, Gaithersburg, MD 20899, USA.}
\affiliation{Department of Physics, Washington University in St. Louis, St. Louis, MO 63130, USA}

\author{ Shanta R. Saha }
\affiliation{Department of Physics, Quantum Materials Center, University of Maryland, College Park, MD 20742, USA.}

\author{Daniel F. Agterberg}
\affiliation{Department of Physics, University of Wisconsin, Milwaukee, Milwaukee, Wisconsin 53201, USA.}

\author{J\"org Schmalian}
\affiliation{Institute for Quantum Materials and Technologies,  Karlsruhe Institute of Technology, Karlsruhe 76021, Germany}

\author{ Nicholas P. Butch}
\affiliation{Department of Physics, Quantum Materials Center, University of Maryland, College Park, MD 20742, USA.}
\affiliation{NIST Center for Neutron Research, National Institute of Standards and Technology, Gaithersburg, MD 20899, USA.}

\author{ Johnpierre Paglione}
\affiliation{Department of Physics, Quantum Materials Center, University of Maryland, College Park, MD 20742, USA.}
\affiliation{NIST Center for Neutron Research, National Institute of Standards and Technology, Gaithersburg, MD 20899, USA.}
\affiliation{The Canadian Institute for Advanced Research, Toronto, Ontario, Canada.}

\author{Aharon Kapitulnik}
\affiliation{Geballe Laboratory for Advanced Materials, Stanford University, Stanford, CA 94305}
\affiliation{Department of Applied Physics, Stanford University, Stanford, CA 94305}
\affiliation{Stanford Institute for Materials and Energy Sciences, SLAC National Accelerator Laboratory, 2575 Sand Hill Road, Menlo Park, CA 94025}
\affiliation{Department of Physics, Stanford University, Stanford, CA 94305}



\date{\today}

\begin{abstract}
Time-reversal symmetry breaking (TRSB) in UTe$_2$ was inferred from observations of a spontaneous Kerr response in the superconducting state after cooling in zero magnetic field, while a finite $c$-axis magnetic field training was further used to determine the nature of the non-unitary composite order-parameter of this material. Here we present an extensive study of the magnetic-field-trained Kerr effect, which unveils a unique critical state of pinned ferromagnetic vortices.  We show that a remanent Kerr signal that appears following the removal of a training magnetic field, which reflects the response of the TRSB order parameter and the external magnetic field through the paramagnetic susceptibility. This unambiguously demonstrate the importance of the ferromagnetic fluctuations and their intimate relation to the composite order parameter.  Focusing the measurement to the center of the sample, we are able to accurately determine the maximum field that is screened by the critical state and the respective critical current. Measurements in the presence of magnetic field show the tendency of the superconductor to produce shielding currents that oppose the increase in vortex-induced magnetization due to the diverging paramagnetic susceptibility.
\end{abstract}


\maketitle


\section{Introduction}

UTe$_2$ is an example of superconductivity in the presence of ferromagnetic fluctuations that persist to temperatures approaching zero, suggesting a magnetic quantum critical phenomenon. The observations of two successive superconducting transitions in specific heat and spontaneous polar Kerr effect (PKE)  \cite{Hayes2020} along the crystallographic $c$-axis led to the conclusion that superconductivity in UTe$_2$ is characterized by a two-component order-parameter that breaks time reversal symmetry \cite{Hayes2020,nevidomskyy2020,Shishidou2020}. While several possibilities were considered for the symmetry representation, the one that seemed to agree with the data was $(\psi_1\in B_{3u},\psi_2\in B_{2u})$ \cite{Hayes2020,Shishidou2020}, implying a non-unitary order-parameter, which we previously argued can be stabilized by magnetic fluctuations \cite{Hayes2020}. 

However, the nature of these magnetic fluctuations is still not fully understood.  Muon spin relaxation indicate coexistence of ferromagnetic fluctuations and superconductivity, where the temperature dependence of the dynamic relaxation rate down to 0.4 K agrees with the self-consistent renormalization theory of spin fluctuations for a three-dimensional weak itinerant ferromagnetic metal \cite{Sundar2019}.  Magnetization measurements can be collapsed onto a single curve using a theory of metallic ferromagnetic quantum criticality \cite{Kirkpatrick2015}, strongly suggesting a zero-temperature ferromagnetic transition. However, recent neutron scattering results show that magnetic fluctuations in UTe$_2$ are dominated by incommensurate spin fluctuations near an antiferromagnetic ordering wave vector and extends to at least 2.6 meV, suggesting that these fluctuations play an important role in the development of the superconducting state \cite{Duan2020}. Furthermore, recent transport and thermodynamics studies under hydrostatic pressure suggest that the superconducting transition temperature is maximized near a putative antiferromagnetic quantum critical point occurring at a modest pressure of $\sim1.3$ GPa \cite{Thomas2020}. While two superconducting states have been confirmed, the nature of the magnetic interaction associated with each of the states could be different \cite{Thomas2020,Thomas2021}. Nevertheless, extrapolation from other similar uranium-based superconductors including URhGe, UCoGe and UGe$_2$ strongly support a spin triplet state mediated by ferromagnetic fluctuations \cite{Hayes2020,nevidomskyy2020,Shishidou2020}. Focusing on the superconducting state at ambient pressure, one way to resolve the magnetic nature of the superconducting state of UTe$_2$ is to test what magnetic state the superconducting order-parameter nucleates upon interaction with the magnetic fluctuations in both, the Meissner and  vortex states.

While the time-reversal symmetry breaking (TRSB) property could be determined unambiguously from the observation of spontaneous Kerr response in the superconducting state after cooling in zero magnetic field, understanding the field-cool data require more attention. First, the strong ferromagnetic fluctuations at low temperatures may counter the onset of Meissner currents. In addition, indications are that UTe$_2$ exhibits very strong pinning effects \cite{Paulsen2021}. For example, removing the magnetic field  at low temperatures after a field-cool may result in a finite remanent magnetization due to trapped flux \cite{Clem1979}, which interacts with the fluctuating magnetic moments as well as with the TRSB superconductor \cite{Hayes2020}. Such a behavior would be enhanced when the material is cooled above the lower critical field, $H_{c1}$, and a critical state concept is expected to be applicable \cite{Bean1962,Bean1964}. In this case Kerr effect will be an ideal probe of the local magnetization, particularly aiming at the center of the sample's surface perpendicular to the applied magnetic field. 

In this paper we report on high resolution polar Kerr effect measurements of single crystal UTe$_2$ samples under various conditions involving combinations of field cool (FC), zero field cool (ZFC), field warmup (FW) and zero field warmup (ZFW). The assemblage of our results suggest strong correlations between the time-reversal symmetry breaking order-parameter, the ferromagnetic susceptibility, and the external magnetic field - when applied.  First, as previously reported \cite{Hayes2020}, the observation of a finite Kerr response under ZFW conditions that followed ZFC, unambiguously prove the time reversal symmetry breaking of the superconducting order-parameter. The varying magnitude of the effect (with a maximum amplitude of $\sim 0.4\mu$rad) and its random sign observed under those conditions is expected from a spontaneous symmetry breaking effect that exhibit two-fold domains.  Furthermore, ZFW measurements that followed a finite FC schedule show a low-temperature, field-independent, Kerr response of $\sim 0.4\mu$rad when the cooling field is $\lesssim 30$G, and an increasing low temperature Kerr response for larger cooling magnetic fields. This remanent Kerr response, which originates from a combination of remanent magnetization associated with trapped vortices, as well as the superconducting order-parameter extrapolates to saturate at $H^*\sim 920$G. The above behavior, as well as the temperature dependence of the remanent Kerr effect, particularly the way it vanishes near $T_c$ suggest an unusual interplay between magnetism and the TRSB order-parameter in a non-unitary superconductor.

Our observation that the TRSB in UTe$_2$ can be trained by a magnetic field along the crystallographic $c$-axis (coinciding with the $z$-direction in our notations) requires the presence of a term $\sim H_z i(\psi_1\psi_2^*-\psi_1^*\psi_2)\equiv H\tilde{\psi}$. Hence, $\tilde{\psi}$ is a TRSB composite order-parameter.  A resulting $z$-axis magnetization will yield the following contribution to the free energy
\begin{equation}
f_m=\alpha m^2 + \gamma m \tilde{\psi} -mH 
\label{interact}
\end{equation}
Since the Curie temperature is suppressed to $T=0$, and to match with the normal state, $2\alpha =\chi_n(T)^{-1}$ is the inverse normal-state magnetic susceptibility. At zero magnetic field the magnetization associated with the order-parameter can be shown to stabilize the nonunitary state \cite{Hayes2020}. However, in the presence of magnetic field, and particularly in the vortex state, the resulting magnetization should reflect a competition between the normal state component and the superconducting tendency to screen the nucleation of a finite magnetization. Here we present a simple analysis of the electromagnetic response of a strongly paramagnetic superconductor, which seems to account for the experimental results that demonstrate the effect of screening in strongly reduce the magnetization as the temperature tends to zero.

\section{Experiment}

To study the interplay between time reversal symmetry breaking (TRSB) in the superconducting state of UTe$_2$, and the vortex state, we performed high resolution polar Kerr effects (PKE) using a Zero-Area Sagnac Interferometer  (ZASI) that probes the sample at a wavelength of 1550 nm, in a He-3 cryostat with base temperature $\lesssim 0.3$ mK.  The UTe$_2$ single crystal used in this study and a basic schematic of the low-temperature setup are shown in Fig.~\ref{zf}, while the full apparatus is described in the supplementary Information (SI) \cite{supplement}.

In general, the Kerr effect is defined through an asymmetry of reflection amplitudes of right- and left- circularly polarized light from a given material, yielding a Kerr rotation angle $\theta_K$, and is observed only if reciprocity is broken. Thus, in the absence of any time-dependence or irreversible effects, a finite Kerr effect unambiguously points to TRSB in that material system.  From its nature, Kerr effect is described within the general theory of scattering, and thus it is ideal to probe TRSB in superconductors since it is not subjected to shielding (Meissner) effects that counter the magnetization. However, restrictions on the possible observation of a finite Kerr effect suggest that for pure crystals, measurements at frequencies where interband pairing effects are important yield a maximum signal, although still very small \cite{Wysokinski2012,Taylor2012,Gradhand2013,Kallin2016}.

In the Kerr measurements reported here we use a collimated beam of diameter  $\sim 10.6~\mu$m, which emerges from a lens and quarter wave plate assembly that were designed to minimize sensitivity to changes in both angle and distance relative to the sample as to prevent temperature-dependent changes in optical alignment \cite{Schemm_Thesis}. For all the measurements the beam is aimed at the center of the sample of average diameter 1.2 mm.  However, probing the system at near-IR frequencies ($\omega$), which is much larger than the superconducting gap energy ($\Delta$) will reduce a typical ferromagnetic-like response of order $\sim1$ rad, by a factor of $(\Delta/\hbar\omega)^2\sim10^{-7}$, yielding a theoretically predicted signal of about 0.1-1 $\mu$rad. Thus, a sensitive device is needed to detect such small signals, and the ZASI with its high degree of common-mode rejection for any reciprocal effects (e.g. linear birefringence, optical activity, etc.) is probably the most suitable one for this study.

In a typical FC Kerr experiment of a weak-pinning superconductor, the sample is first cooled in a field (FC) lower than $H_{c1}(0)$. The Kerr effect is then measured at zero-field while warming up (ZFW). In this case the ZFW measurement yields results similar to the largest Kerr value obtained in a zero-field cool (ZFC) followed by ZFW experiment, indicating a single domain Kerr response. If the sample is cooled in a field larger than $H_{c1}(0)$, some flux will be trapped after removing the magnetic field at low temperatures, which, depending on the magnetic susceptibility of the material, may result in a finite Kerr effect. Determining the magnetic contribution above $T_c$, we expect that upon cooling in a field $H_{c1}\lesssim H\ll H_{c2}$, a normal state contribution will be smaller by at least a factor $H/H_{c2}$. Thus, without explicit magnetism (that is, magnetic response beyond Pauli paramagnetism or Landau Diamagnetism), such effects are undetectable. Indeed,  in previous Kerr measurements of Sr$_2$RuO$_4$ \cite{Xia2006}, where Kerr effect of $\sim 0.1\mu$rad was detected, or the heavy fermion Uranium-based superconductors UPt$_3$ \cite{Schemm2014} and URu$_2$Si$_2$ \cite{Schemm2015}, and the filled-skutterdite PrOs$_4$Sb$_{12}$ \cite{Levenson2018} (giving larger signal of $\sim 0.4$ to 0.7 $\mu$rad, which is expected due to their strong spin-orbit interaction), the same magnitude of Kerr effect was observed irrespective of the magnitude off the orienting FC.  Testing the apparatus with reciprocal reflecting media such as simple BCS superconductors (or just mirrors),  as well as the spin-singlet d-wave heavy-fermion compound CeCoIn$_5$ yielded a null result  as expected \cite{Schemm2017}.  

\begin{figure}[ht]
\centering
\includegraphics[width=1.0\columnwidth]{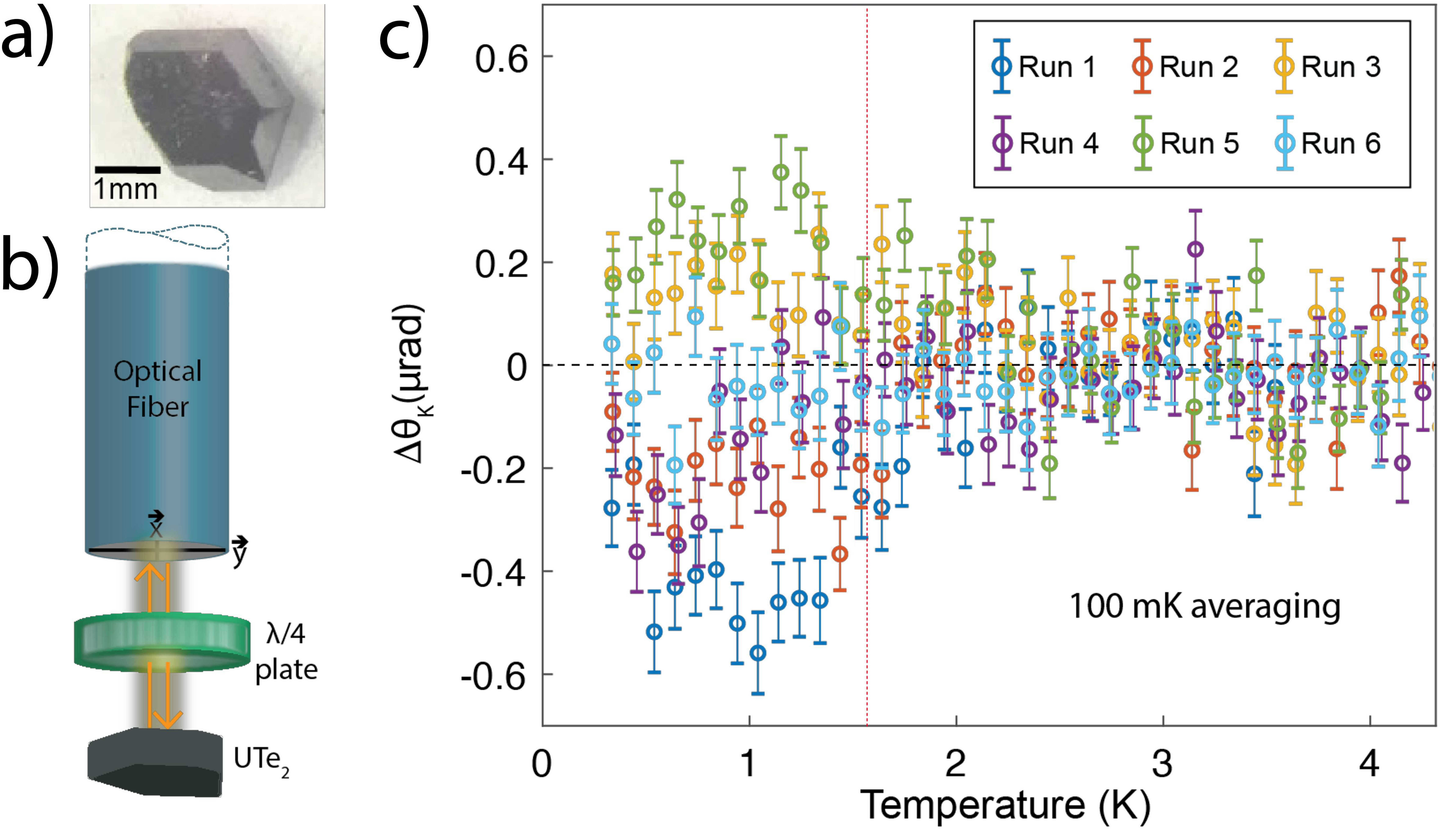} 
\caption{(a) Crystal used for Kerr measurements. (b) End part of the apparatus (for full description see SI \cite{supplement}). (c) Five different Zero-field-cool Kerr data for a  superconducting UTe$_2$ with $T_c\approx 1.55$K }
\label{zf}
\end{figure}

\section{Results}

{\it Zero field data:} Polar Kerr effect measurements were performed at low temperatures on a single crystal of UTe$_2$, after first cooling it to a base temperature of 0.3 K in ambient field of $\sim0.2$ G (which we will denote as zero field) and measuring while warming the sample to above the superconducting transition of $T_c\approx 1.55$ K. The cumulative results of five cooldowns are shown in Fig.~\ref{zf}c, and are consistent with measurements on a second crystal. We note that different cooldown result in a random size and random sign of the signal, bound by a maximum signal of $|\theta_K|\approx 0.4\mu$rad at $0.3$ K. While these zero-field cool data may point to domain formation, it is typically assumed that with a two-state order-parameter, as we anticipate in UTe$_2$, domains may be costly, thus often measurements will result in the full signal.\\

{\it Low field data:} Assuming the zero-field data arises from the TRSB order-parameter, we use field cool data, with the field oriented along the $c$-direction of the crystal, to test for the ability to couple to that order-parameter, and at the same time orient the sample to a single domain. This is a crucial part of the correct determination of the possible components of the proposed nonunitary order-parameter \cite{Hayes2020}.
\begin{figure}[h]
	\centering
	\includegraphics[width=1.0\columnwidth]{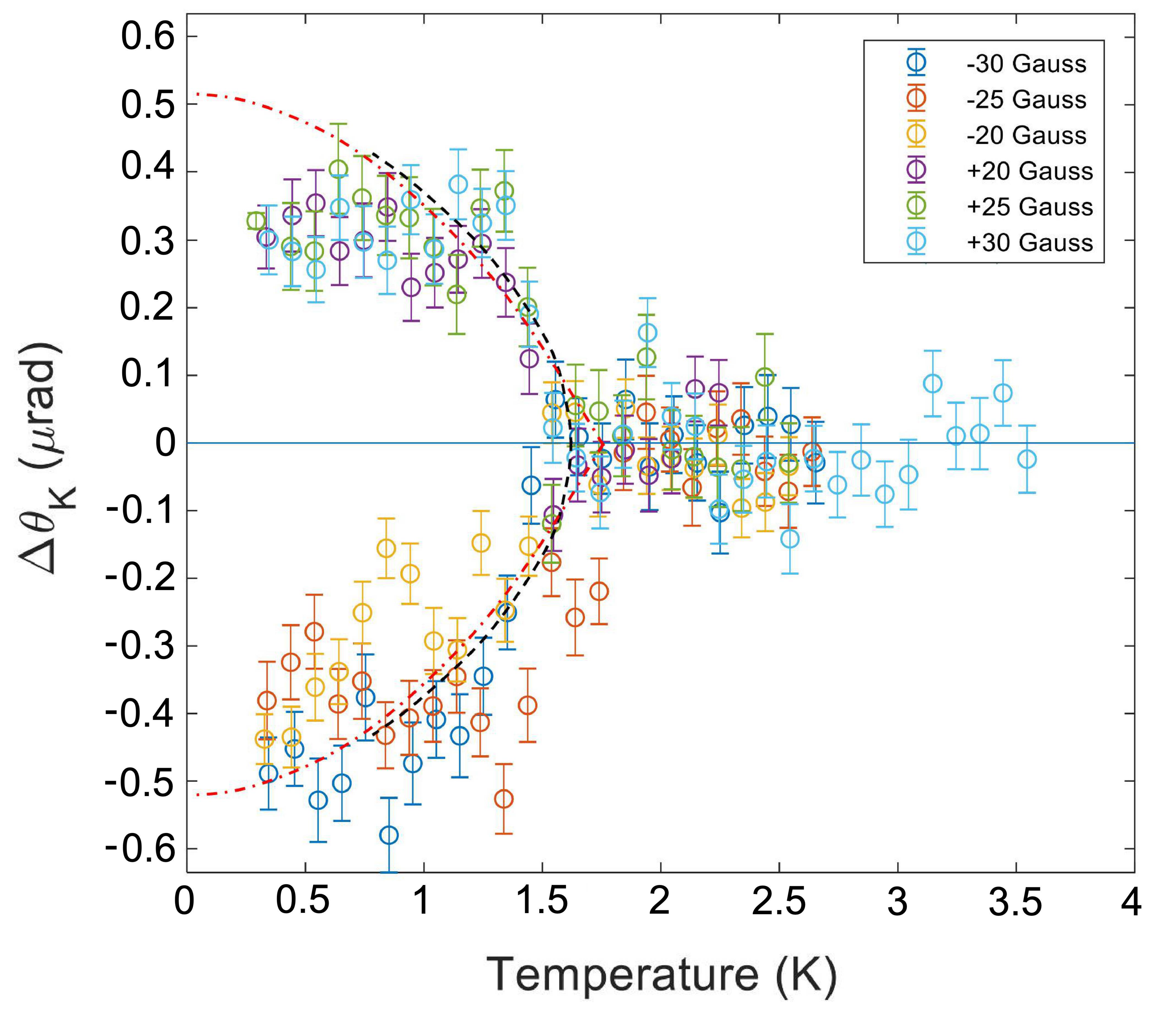}
	\caption{Very Low Field Cool - Measurements at zero-field Warmup. Note that the saturated Kerr angle is the same for this field range. Here dashed-dotted line (red) is a fit to $[1-(T/T_c)^2]$ with $T_c=1.7$ K, while dashed line (black) is a fit to a $[1-(T/T_c)]^{1/2}$ with $T_c=1.55$ K, which should dominate near the actual TRSB temperature.}
		\label{vlowfield}
\end{figure}

First we note that for FC measurements of $H\lesssim 15$G we see no difference from ZFC experiments, that is, values can fluctuate in each cooldown similar to Fig.~\ref{zf}. Figure~\ref{vlowfield} shows a set of 6 different experiments where the sample was cooled in magnetic fields of $\pm20$, $\pm25$, and $\pm30$G, and the Kerr angle was measured in a ZFW configuration. The low-temperature Kerr angle of this collection of data seems to saturate at the same value, which also coincides with the maximum value obtained in the ZFC experiments. We therefore conclude that this value of $\theta_K \approx 0.4 ~\mu$rad is the intrinsic contribution due to the TRSB order-parameter. Attempting to fit the temperature behavior to the standard form of an order parameter $\propto[1-(T/T_c)^2]$ yields a reasonable, but with a slightly higher $T_c$ than where the Kerr effect seems to vanish.   A better fit $\propto [1-(T/T_c)]^{1/2}$ with the actual $T_c$ seems to hold near where the Kerr effect vanishes, which should dominate near the actual TRSB temperature, thus indicating the existence of a composite order parameter.  However, below $\sim T_{c}/2$, $\theta_K(T)$ develops an anomalous behavior, exhibiting more flat temperature dependence.  Such a behavior may be a precursor to a more pronounced flat temperature behavior that we observe upon cooling in higher magnetic fields, and is typical for remanent magnetization behavior of a ``hard superconductor,'' which exhibits large vortex critical current \cite{Clem1979,Clem1993}. Its occurrence in UTe$_2$  is a central theme of this manuscript.
\begin{figure}[h]
	\centering
	\includegraphics[width=1.0\columnwidth]{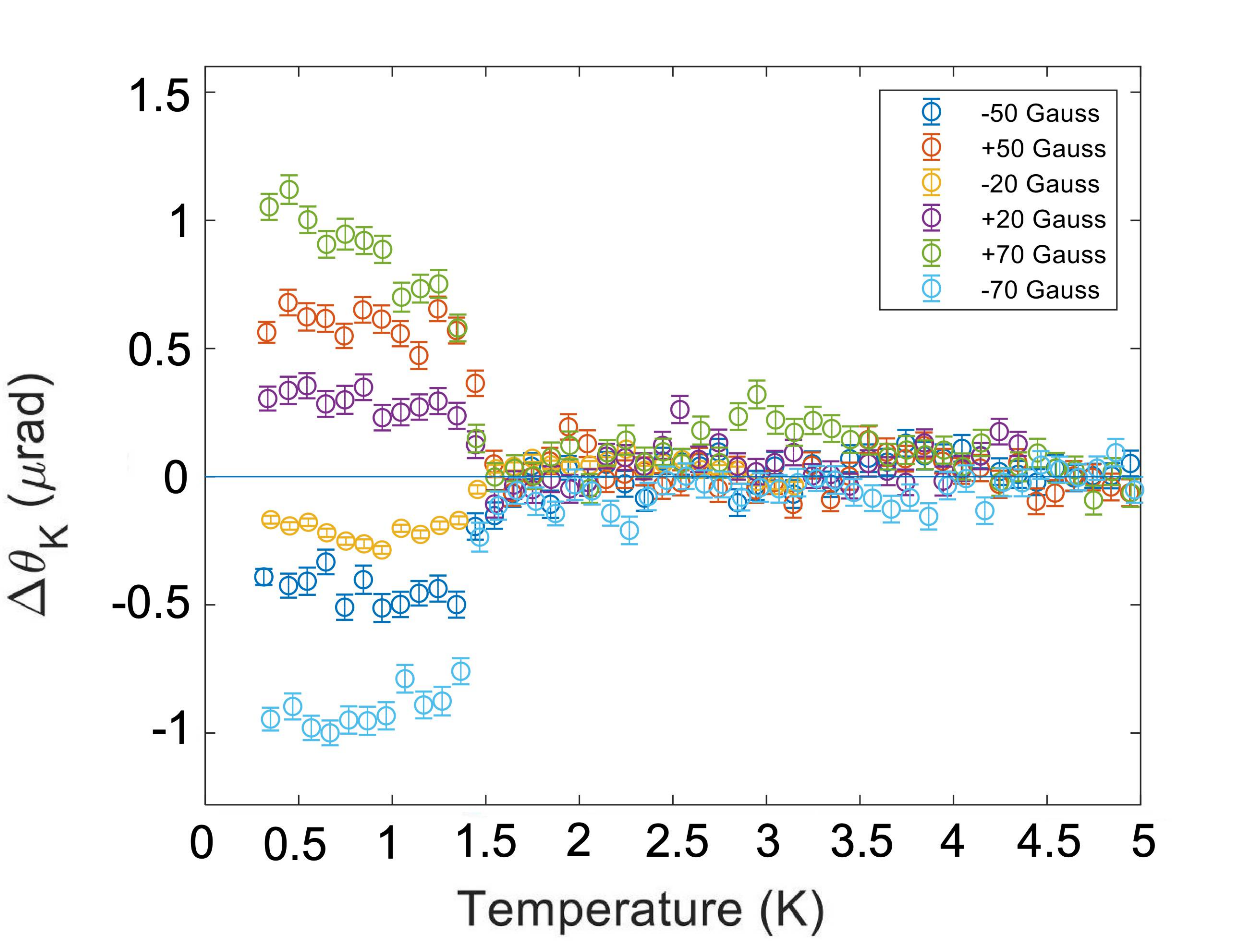}
	\caption{Low Field Cool. dashed line box is the low field regime shown in Fig.~\ref{lowfield}.}
		\label{lowfield}
\end{figure}

Increasing the training magnetic field above $\sim30$G, Fig.~\ref{lowfield} shows that the low-temperature Kerr angle saturation value now increases with field, and continues to show a rather flat behavior in approaching $T_c$, thus further support our hypothesis that the Kerr response through $\sigma_{xy}(\omega)$ originates from the intrinsic magnetism. However, the increase in remanent Kerr value indicates that the vortex state induced by the external cooling field now dominates, which is observed through the interaction with the induced ferromagnetism.\\

\begin{figure}[h]
	\centering
	\includegraphics[width=1.0\columnwidth]{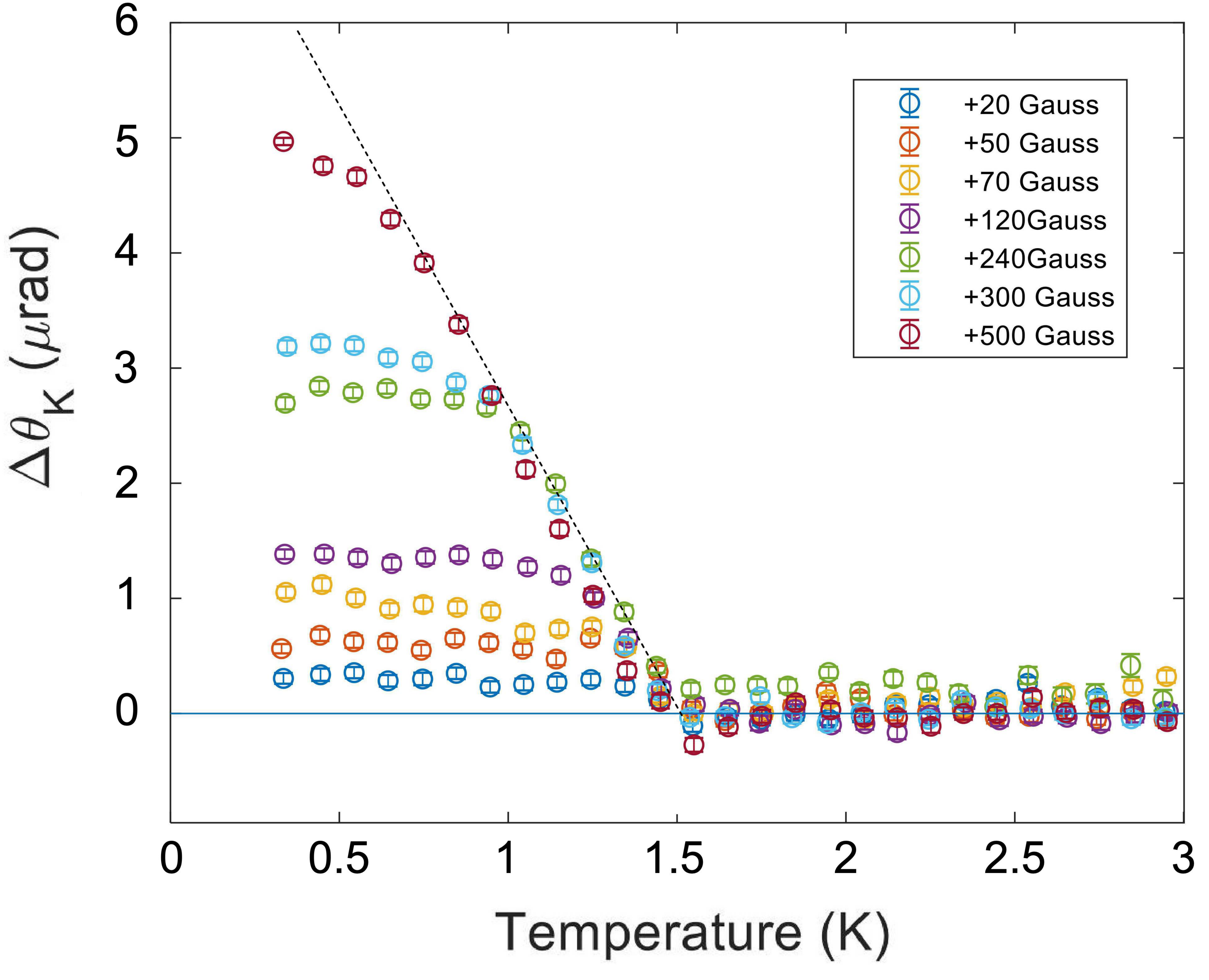}
	\caption{Temperature dependence remanent Kerr effect after Field Cool. Dashed line denotes the ``envelope'' behavior as the cooling field is increased. It extrapolates to $\Delta \theta_K \approx 8.0\mu$rad (dashed line), which would correspond to an approximate cooling field of  $\sim 920$G using extrapolation of the fit shown in Fig.~\ref{slope}}
		\label{fields}
\end{figure}

{\it Wider field range:} When further increasing the magnetic field, a closer resemblance to the behavior of the magnetization of a strong-pinning superconductor in the critical state is revealed. For example, inspection of remanent magnetization in field cooled measurements on URu$_2$Si$_2$ and UPt$_3$ \cite{Koziol1994} show an evolution very similar to our Kerr measurements in Fig.~\ref{lowfield} and the subsequent increased field regime shown in Fig.~\ref{fields}. At the same time, we notice that in our previous measurements of Kerr effect in this materials the TRSB Kerr signal was independent of the training field \cite{Schemm2014,Schemm2015}, thus corroborating the claim that the Kerr effect in these materials detects directly the TRSB order-parameter.
\bigskip

\begin{figure}[ht]
\centering
	\includegraphics[width=1.0\columnwidth]{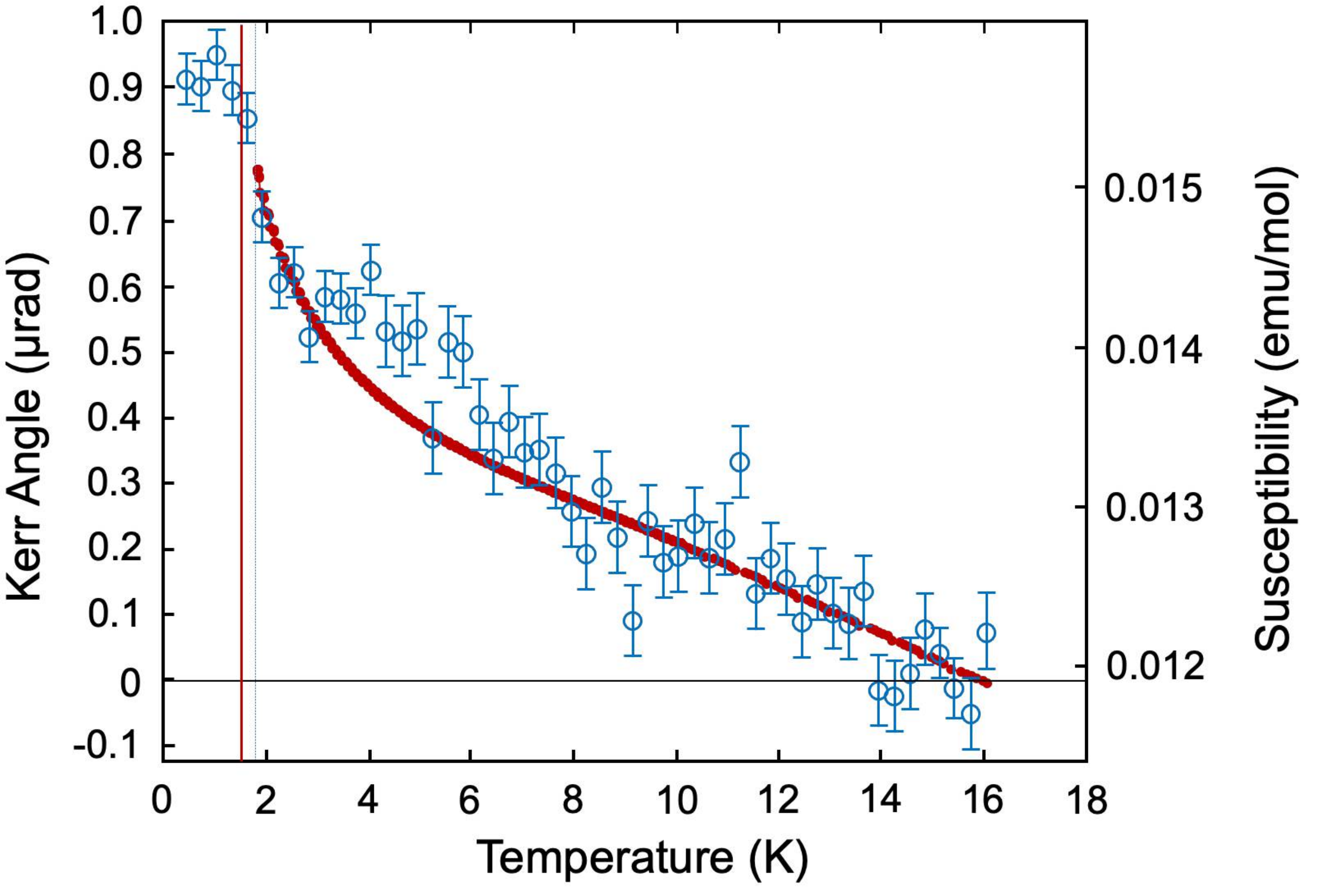}
		\caption{ Susceptibility and Kerr effect measurements of UTe$_2$ in magnetic fields below $H^*$. For Kerr effect he sample was cooled in magnetic field of 240G and then measured when warmed up in 240G (data points represent bin-average of 0.3 K).  Susceptibility data was measured in a SQUID magnetometer down to 1.8 K, and was fitted to the Kerr data in the temperature range of 7 to 8 K assuming $\theta_K \propto \chi=M/H$.  The vertical line is the zero-field warmup Tc. }
	\label{infield}
\end{figure}

{\it Measurements in finite magnetic field:}
The conclusion in \cite{Hayes2020} that the order-parameter is non-unitary with $(\psi_1\in B_{3u},\psi_2\in B_{2u})$, which is stabilized by magnetic fluctuations, suggests the possibility that in the presence of magnetic field the composite order-parameter may appear at a higher temperature, either at the higher $T_c$, or as fluctuation induced effect \cite{Schmalian}.  Assuming a standard correspondence between the Kerr and magnetization measurements (see e.g. \cite{Xia2009}), Fig.~\ref{infield} shows a comparison of Kerr effect measurement below 16K and SQUID magnetometry of a similar sample from the same batch. Measurements were done at 240G along the $c$-axis (below $\sim 1000$G the magnetic susceptibility is roughly field independent, possibly exhibiting a weak minimum at $\sim 240$ G. See SI \cite{supplement}). Also marked in Fig.~\ref{infield} is the $T_c$ as determined from Fig.~\ref{lowfield}. While there is a good correspondence between the Kerr response and the magnetization, an anomaly in the Kerr effect is detected in the 2 to $\sim$ 6 K, which could be traced to the interplay between the magnetization along different axes in that regime (see SI \cite{supplement}). 

\section{Analysis and Discussion}

Magneto-optical effects are described within quantum theory as the interaction of photons with electron spins through spin-orbit interaction (see e.g.~\cite{Pershan1967}). In a ferromagnetic material the Kerr angle is found to scale with the magnetization \cite{Argyres1955}, while inherently antiferromagnetic systems may still exhibit a finite Kerr effect depending on their local symmetry (see e.g. \cite{Orenstein2011}). 

Focusing on the measurements on UTe$_2$, we first note that for all FC measurements of $H\lesssim 15$ G, including ZFC, the sample exhibits a random in sign and in magnitude Kerr signal, which is bound by $|\theta_K(0)|\approx 0.4~\mu$rad. This clearly establishes an intrinsic TRSB associated with the superconducting order-parameter rather than a mere magnetic response.  However, as will be discussed below, UTe$_2$ is a strong-pinning superconductor, clearly exhibiting critical state features, which will require further considerations in analyzing the low field data.\\

{\it The low-field data:} The fact that no field dependence is observed for ZFW measurements that follows either ZFC or FC of $H\lesssim 30$ G, suggests that this very low field effect is a result of either the TRSB order-parameter, or a constant remanent magnetization, which is induced by an internal ordering field rather than the external cooling field. This may occur if the cooling field $H$ aligns the order-parameter $\tilde{\psi}$, which in turn induces a finite magnetization through the bilinear coupling term as in Eqn.~\ref{interact}. Indeed, Paulsen {\it et al.} \cite{Paulsen2021} reported that expulsion of flux in FC experiments was negligible in fields greater than just a few G. Such bulk magnetization measurements may reflect a critical state that persists even below $H_{c1}$ \cite{Krusin-Elbaum1990,Clem1993}. By contrast, in our Kerr measurements the $\sim 10.6~ \mu$m diameter beam points at the center of the sample of average diameter 1.2 mm (see Fig.~\ref{zf}b). Thus, in such measurement along the applied magnetic fields, we expect the center of the sample to practically be vortex free (In the pure sense of the critical state model it is completely vortex free).\\

\begin{figure}[h]
	\centering
	\includegraphics[width=0.9\columnwidth]{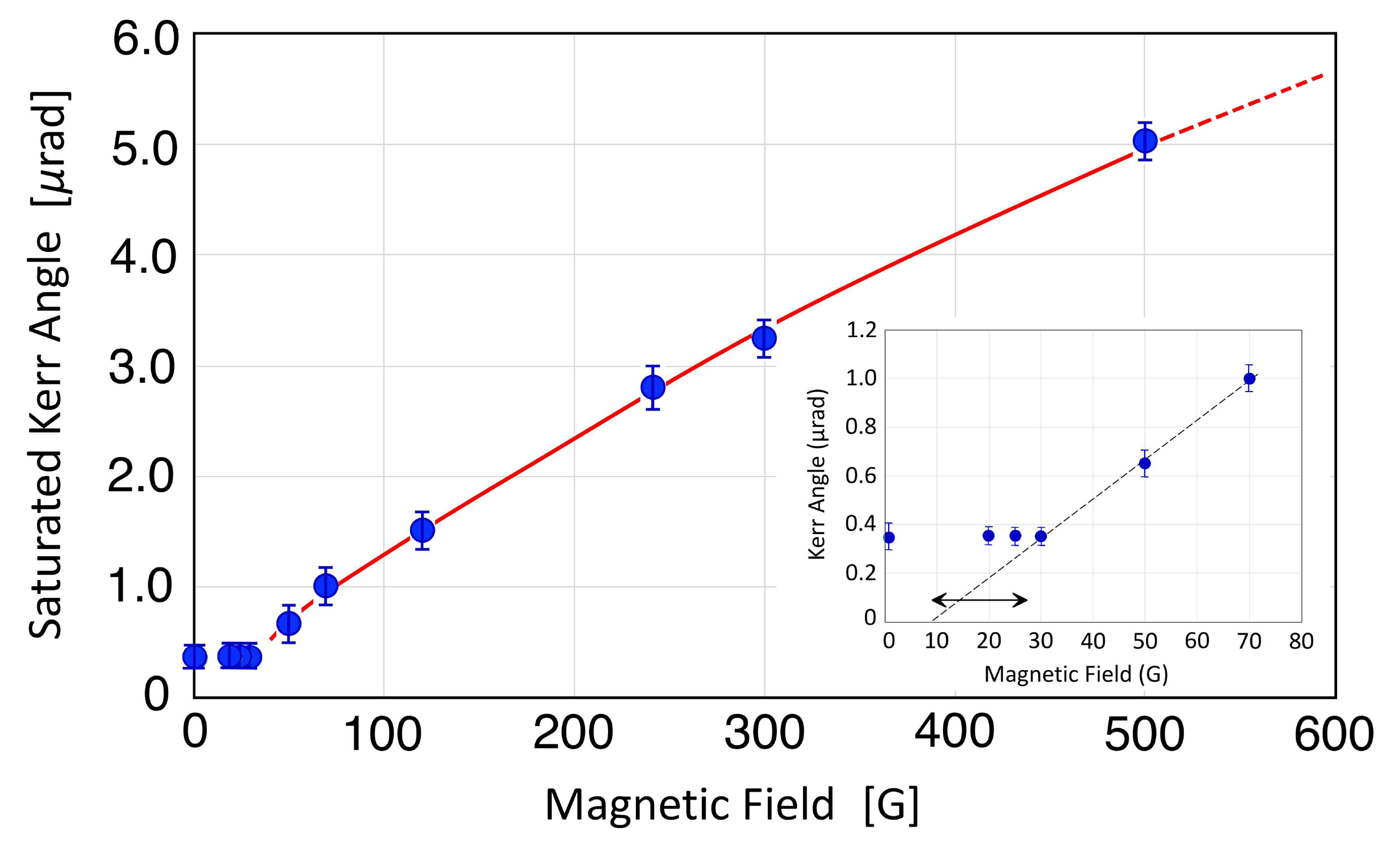}
	\caption{saturated Kerr vs. field UTe$_2$}
		\label{slope}
\end{figure}

{\it Flux pinning, the critical state and high field data:} UTe$_2$ is a strongly paramagnetic material, with large magnetic susceptibility above the superconducting $T_c$. Further observations of strong magnetic hysteresis, weak Meissner effect, and large critical current \cite{Paulsen2021} imply the presence of a critical state \cite{Clem1979}, where cooling in magnetic field may result in large remanent magnetization. While this may impact field-cool Kerr measurements, the configuration of the experiment may also provide new insight into the critical state phenomenon in this material. Moreover, it was shown by Clem and Hao \cite{Clem1993} that with strong pinning, even below $H_{c1}(0)$, cooling in a field results in remanent magnetization that is only weakly temperature dependent when warming up, and disappears only above the irreversibility temperature $T_{irr}<T_c$ \cite{Koziol1994}. Increasing further the magnetic field will result in an increase in the remanent magnetization, until a field $H^*(T)$ is reached, where for a given temperature $T$, full flux penetration into the center of the sample occurs.   Assuming that in this field regime the Kerr angle is proportional to the remanent magnetization in the material, we first extrapolate the envelope of the temperature dependence of the $\theta_K(T)$ to $T=0$, obtaining an extrapolated value of $8.0\pm0.3 \mu$rad. Next we plot the saturated Kerr angle from Fig.~\ref{fields} as a function of cooling field as shown in Fig.~\ref{slope}. Combining these two measurements we find that the Kerr angle at $T=0$ will be maximum in a field-cool of $H_{max}\approx 920$G. 
Since we assume that the vortex state, which induces the ferromagnetic magnetization, is in the critical state, the vortex distribution in space is determined by the critical current which is assumed constant \cite{Bean1962}. In this model the vortex-density profile (i.e. the profile of the magnetic induction $B(r)$) is a straight line, matching the external field at the sample surface. If the external field is taken to zero before warming up, this profile determines the variation of the remanent magnetization in space. Obviously, with the above approximation the maximum remanent magnetization is obtained for a magnetic field $H^*$ for which a first vortex is introduced to the center of the sample. Similarly, we will need to take the external magnetic field to $-H^*$ to subsequently have the center of the sample completely screened. For a sample of averaged diameter $D$, the above arguments imply:
\be
H^*=\frac{2\pi J_c D}{c}, \ \ \ \ \  {\rm or} \ \ \ \ \ \  J_c=\frac{c}{2\pi D}H^*
\ee
Taking for our data $H^*\approx H_{max}=920 G$, we obtain $J_c\approx 1.05\times 10^4$A/cm$^2$, which is very close to the critical currents recently measured for UTe$_2$ of $J_c^a\approx 0.8\times 10^4$A/cm$^2$ and $J_c^b\approx 1.6\times 10^4$A/cm$^2$ \cite{Paulsen2021}. This agreement between the Kerr data and the direct measurements of the critical current further support our interpretation of the origin of the Kerr signal above $H_{c1}$.\\
\begin{figure}
\includegraphics[width=0.9\columnwidth]{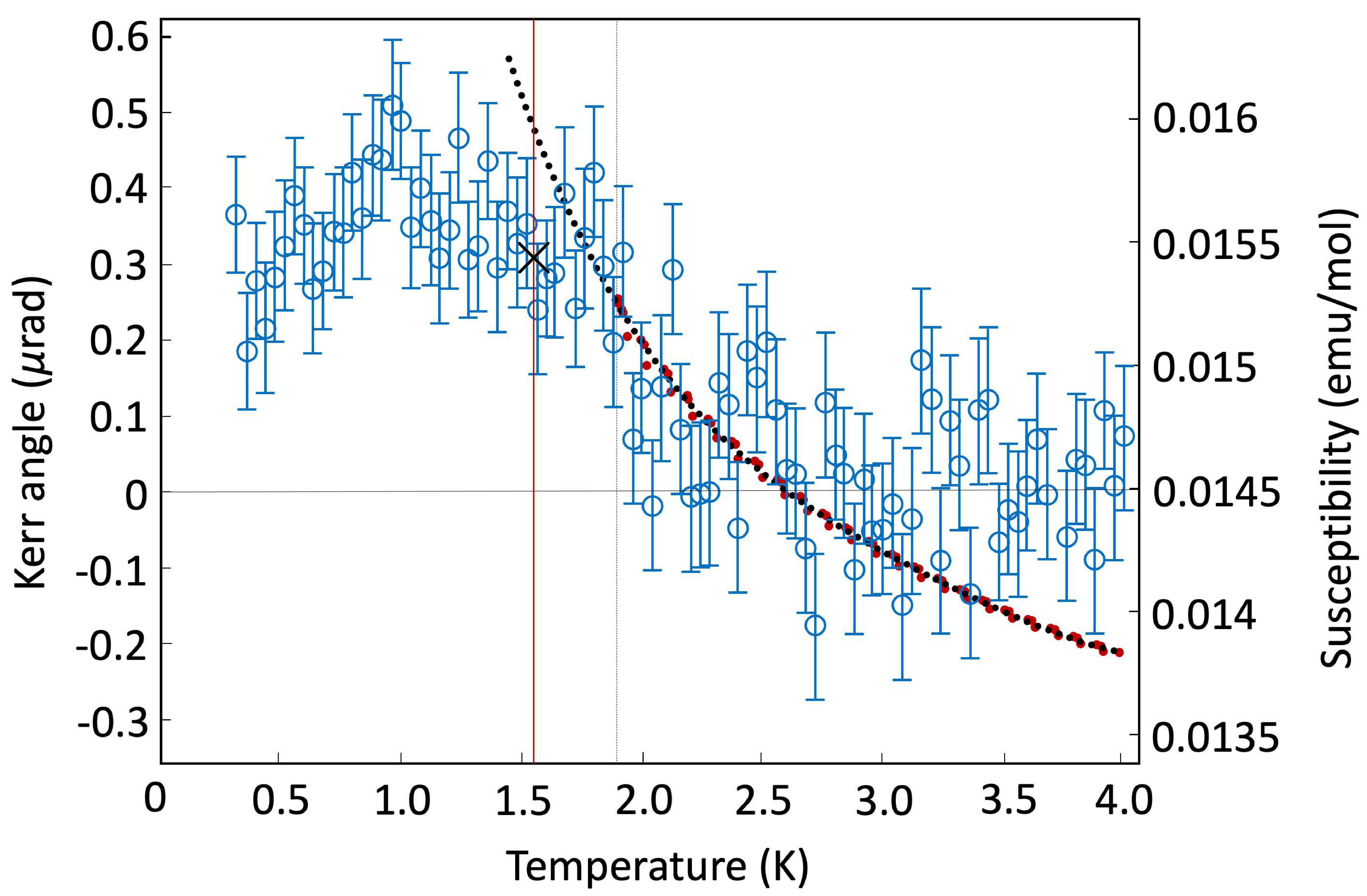}
	\caption{ Expanded vertical scale of field cool, field warmup of Fig.~\ref{infield}(b). Zero Kerr angle was offsetted  to the plateau around 4K.  The SQUID susceptibility data (red full circles) is the same as in Fig.~\ref{infield} for this temperature range. The dotted line is the extrapolation of the SQUID-measured susceptibility shown in Fig.~\ref{infield} fitted below 4K to the lowest susceptibility temperature measurement of 1.8K (represented with a faint vertical line). The vertical red line is the zero-field warmup $T_c =1.55$K, which meets the local averaged Kerr data at the $\times$ symbol.}
		\label{infieldTc}
\end{figure}

Further support of the critical state scenario is discussed in the SI: \cite{supplement}), where we analyze a hysteresis loop at a field much smaller than $H^*$, demonstrating how the evolution of the superconducting order counter-acts the increase in magnetization below $T_c$. While cooling in a field nucleates remanent magnetization, cooling at zero field and increasing the field to $H_{c1}\ll H < H^*$ results in a weak Kerr effect at the center of the sample, which is further reduced almost completely as the field is reduced to zero.

It is important to note the difference between UTe$_2$ and other uranium-based superconductors that exhibit a critical state. Focusing on UPt$_3$ and URu$_2$Si$_2$, Kozio\l{} {\it et al.}, \cite{Koziol1994} show remanent magnetization curves for these two materials for field below their respective $H^*$.  At the same time, our own Kerr measurements on these materials \cite{Schemm2014,Schemm2015} show that the Kerr effect in FC measurements is independent of cooling field, and coincide with the envelope of the zero-field cool experiments. The reason is, as explained above, that with no explicit magnetism, the trapped vortices only carry a Kerr effect due to Pauli paramagnetism or Landau diamagnetism, which is reduced by those vortices' relative density of $H/H_{c2}$. Such an effect is undetectable with our nanoradians resolution. At the same time, the Kerr measurements on UTe$_2$ do resemble the bulk magnetization measurements of UPt$_3$ and URu$_2$Si$_2$ \cite{Koziol1994} (see SI: \cite{supplement}). This clearly indicate that the core of the vortices in UTe$_2$ are ferromagnetic, with the core magnetization aligned with the direction of the cooling field. This then confirms the tendency of the material towards ferromagnetism and further suggests that ferromagnetic fluctuations affect the superconducting order-parameter.\\

{\it TRSB order-parameter in magnetic field:}  Inspection of Fig.~\ref{infield}, the normal state (FW) Kerr data shows a standard increase consistent with a paramagnetic susceptibility of the material, before experiencing a sharp increase below $\sim6$K which seems to return to the original susceptibility curve below $\sim 2$K.  Indeed, a change in course of the magnetic susceptibility was previously noted in this temperature range by several groups \cite{Ran2019,Ikeda2006}, typically attributed to the $c$-direction being perpendicular to the easy axis-$a$ (See also SI: \cite{supplement}). To better understand the observed behavior, we show in Fig.~\ref{infieldTc} the expanded data below 4K (data points represent bin-average of 100 mK), which includes the superconducting transition region. Since susceptibility data is only available down to 1.8K, our analysis needs to rely on the extrapolation of this normal state data to lower temperatures, however, inspection of the trend of the data already suggests that the extrapolated susceptibility will go above the average of Kerr angle at $T_c$, marked with the cross ($\times$) symbol at the $T_c$ of the ZFW measurements (e.g. Fig.~\ref{fields}). To illustrate this claim, we fit the susceptibility data below 4K (4$^th$ order polynomial give the best fit), and extrapolate it lower than the measured 1.8 K.  The extrapolated susceptibility, which represents the normal-state magnetic response, clearly overshoots above the actual Kerr data. This behavior indicate the emergence of superconducting fluctuations.

Since our experiment does not detect the Meissner currents that shield the whole crystal, the data must indicate the competition between the paramagnetic susceptibility and the superconducting order-parameter that develops in the interior of the sample. This in turn suggests that the proximity to the superconducting state is felt already $\sim 0.3$ K above the Kerr-detected-$T_c$. However, as $T_c$ is approached, the magnetic response weakens and the Kerr angle at the center of the sample decreases towards its value at $T_c$. Such an effect cannot be described by Eqn.~\ref{interact}, which considers a uniform magnetic field and ignores screening effects.  Instead, to understand this effect we need to include the full elecrodynamics of the superconductor and its interaction with the magnetization. Since the Kerr effect is expected to follow the magnetization, including the TRSB effect of the superconducting order-parameter, we revise Eqn.~\ref{interact}, writing
\begin{equation}
f_m=\alpha m^2 + c|\nabla m|^2 +\gamma m \tilde{\psi} -mh
\label{FE}
\end{equation}
where we added a term that allows for non-uniform magnetization, which is expected in the vortex state and the external magnetic field is replaced with the local magnetic field, $\vec{h}(\vec{r})$, that we will take as pointing in the $c$-direction (thus omit the vector sign.) This local field arises from the solution of the Ginzburg-Landau equations and thus its profile through the vortices already include the field penetrating through the core and the shielding currents that extend a penetration-depth distance $\lambda$ beyond the vortex core. Moreover, we recall that $\langle \vec{h}(\vec{r})\rangle = \vec{B}$, the magnetic induction of the superconductor. A more detailed approach, aimed at analyzing the magnetic structure in a ferromagnetic superconductor was recently presented by Devizorova {\it et al.}\cite{Devizorova2019}. However, here we were only interested in the magnitude of the average magnetization.  Thus, to obtain the equilibrium contribution of the paramagnetism to the magnetization in the superconductor, we minimize the magnetic free energy with respect to $m$, obtaining:
\begin{equation}
\frac{1}{\chi_n}\left[1-a^2\nabla^2\right]m=h-\gamma\tilde{\psi}
\label{effective}
\end{equation}
where, as in Eqn.~\ref{interact} we identify $2\alpha =\chi_n(T)^{-1}$ as the inverse of the normal-state magnetic susceptibility, and $a^2=c/\alpha$ is the magnetic stiffness parameter, which represents a microscopic length of order lattice constant. Now, $(h-\gamma\tilde{\psi})$ acts as an effective local magnetic field that follow the vortex lattice periodicity. As we solve for the magnetization and substitute back in Eqn.~\ref{FE}, we clearly observe that the large magnetic susceptibility caused by ferromagnetic fluctuations further enhances the effective field $\sim\chi h$ that couples to the composite order parameter.  Taking the Fourier transform of Eqn.~\ref{effective}, we arrive at the $q$-component of the transform for the magnetization:
\begin{equation}
m_q=\chi_n\frac{h_q-\gamma\tilde{\psi}_q}{1+a^2q^2}
\end{equation}
However, since $h_q$ arises from the London equation, it should decay with the penetration length $\lambda$ away from the vortex center, thus  $h_q=h(0)/(1+\lambda^2q^2)$. We thus expect
\begin{equation}
m_q=\frac{\chi_n}{1+a^2q^2}\frac{h(0)-\gamma\langle \tilde{\psi}\rangle}{1+\lambda^2q^2}
\label{screen1}
\end{equation}
The total magnetization density is then
\begin{equation}
\langle m\rangle = \sum_q\frac{\chi_n}{1+a^2q^2}\frac{B-\gamma\langle \tilde{\psi}\rangle}{1+\lambda^2q^2}
\label{screen2}
\end{equation}
Near $T_c$ the sum is dominated by $q\to 0$, yielding the same result we would expect from Eqn.~\ref{interact}, that is:
\begin{equation}
\langle m\rangle_{T\approx T_c} =\chi_n\left(B-\gamma\langle \tilde{\psi}\rangle\right)
\end{equation}
However, as the temperature tends towards zero, we cannot ignore the shorter wavelengths, which will be  dominated by the vortex lattice wave-vector, $q\approx \sqrt{\Phi_0/B}$. Since in that regime $a^2B/\Phi_0\ll1$, while $\lambda^2B/\Phi_0\approx B/H_{c1}\gg1$, we can change the sum to integral, and (up to order of ${\rm ln}\kappa$) we obtain:
\begin{equation}
\langle m\rangle_{T\ll T_c} \approx \chi_n\frac{H_{c1}}{B}\left(B-\gamma\langle \tilde{\psi}\rangle\right) 
\end{equation}
which demonstrates the effect of screening. Here $\chi_n(T) \to \infty$ as $T\to 0$, before saturating to its full value at $T=0$. Thus without the large reduction of $H_{c1}/B$, the magnetization would monotonically increase towards zero temperature.
Our experimental data for Kerr effect at 240 G, which is much larger than $H_{c1}$ indeed agree with the trend calculated above. This unique behavior is clearly demonstrated in Fig.~\ref{infieldTc}.

\section{Conclusions}

UTe$_2$ exhibits strong paramagnetic fluctuations, with no apparent finite-temperature magnetic order. Using magneto-optic polar Kerr effect to study the magnetic behavior of UTe$_2$ crystals, while focusing our detection beams to the center of the sample, away from any surface Meissner currents, we are able to study the pure superconducting wavefunction response at very low field, and the evolution of the vortex critical state at higher field. This approach allows for a clear demonstration of the properties of a critical state that appears in this material in the presence of an applied magnetic field, including the determination of the critical current density for vortex motion when the field is applied in the $c$-direction. When vortices are present, a ferromagnetic magnetization is detected, confirming the proximity to a ferromagnetic critical point, possibly a quantum-critical one. Subtracting out the magnetic effect, the order-parameter shows an orbital TRSB effect of size similar to the B-phase of UPt$_3$ \cite{Schemm2014} and the low temperature behavior of the superconducting state of URu$_2$Si$_2$ \cite{Schemm2015}. Finally, measurements in the presence of magnetic field clearly show the tendency of the superconductor to produce shielding currents that oppose the increase in vortex-induced magnetization due to the diverging paramagnetic susceptibility.\\

\noindent {\bf Acknowledgments:} We acknowledge discussions with Steve Kivelson and Ronny Thomale. Work at Stanford University was supported by the Department of Energy, Office of Basic Energy Sciences, under contract no. DE-AC02-76SF00515. Work at University of Wisconsin was supported by the Department of Energy, Office of Basic Energy Sciences, Division of Materials Sciences and Engineering under Award DE-SC0021971.  Research at the University of Maryland was supported by the Air Force Office of Scientific Research Award No. FA9550-14-1-0332 (support of T.M.), the Department of Energy Award No. DE-SC-0019154 (specific heat experiments), the Gordon and Betty Moore Foundation's EPiQS Initiative through Grant No. GBMF9071 (materials synthesis), NIST, and the Maryland Quantum Materials Center. Sample synthesis and preliminary characterization were supported by NIST. Work at Karlsruhe was supported by the Deutsche Forschungsgemeinschaft (German Research Foundation) Project ER 463/14-1.

\bibliography{ute2magnetic}

\newpage

\onecolumngrid
\newpage
\setcounter{section}{0}
\setcounter{figure}{0}
\renewcommand{\thefigure}{S\arabic{figure}}
\renewcommand{\theequation}{S.\arabic{equation}}
\renewcommand{\thetable}{S\arabic{table}}
\renewcommand{\thesection}{S\arabic{section}}

\renewcommand{\thefootnote}{\fnsymbol{footnote}}

\begin{center}
\textbf{ SUPPLEMENTARY INFORMATION}

\vspace{3em}
\textbf{Interplay between magnetism and superconductivity in UTe$_2$}\\

\fontsize{9}{12}\selectfont

\vspace{3em}
D.S. Wei,$^{1,2,3}$, D. Saykin,$^{1,3,4}$  O. Y. Miller,$^{1,4}$ D. F. Agterberg,$^5$ N. P. Butch,$^{6,7}$ J. Paglione,$^{6,7,8}$ and Aharon Kapitulnik,$^{1, 2, 3, 4}$
\vspace{1em}
$^1${\it Geballe Laboratory for Advanced Materials, Stanford University, Stanford, CA 94305}\\
$^2${\it Department of Applied Physics, Stanford University, Stanford, CA 94305}\\
$^3${\it Stanford Institute for Materials and Energy Sciences, SLAC National Accelerator Laboratory, \\ 2575 Sand Hill Road, Menlo Park, CA 94025}\\
$^4${\it Department of Physics, Stanford University, Stanford, CA 94305}\\
$^5${\it Department of Physics, University of Wisconsin, Milwaukee, Milwaukee, Wisconsin 53201, USA.}\\
$^6${\it Department of Physics, Quantum Materials Center, University of Maryland, College Park, MD 20742, USA}\\
$^7${\it NIST Center for Neutron Research, National Institute of Standards and Technology, Gaithersburg, MD 20899, USA.}\\
$^8${\it The Canadian Institute for Advanced Research, Toronto, Ontario, Canada.}\\

\end{center}

\vspace{20em}

\section*{ MATERIALS, METHODS AND ADDITIONAL INFORMATION}


\subsection*{Single Crystals Growth}
The UTe2 single crystals were grown by a chemical vapor transport method. Powder x-ray diffraction (XRD), neutron scattering and Laue XRD measurements indicate that the single crystals are of high quality. The details of the sample growth and characterization are given in Refs~\cite{Ran2019,Hutanu2019}.

\subsection*{Measurement Apparatus}
\begin{figure}[h]
\includegraphics[width=0.75\columnwidth]{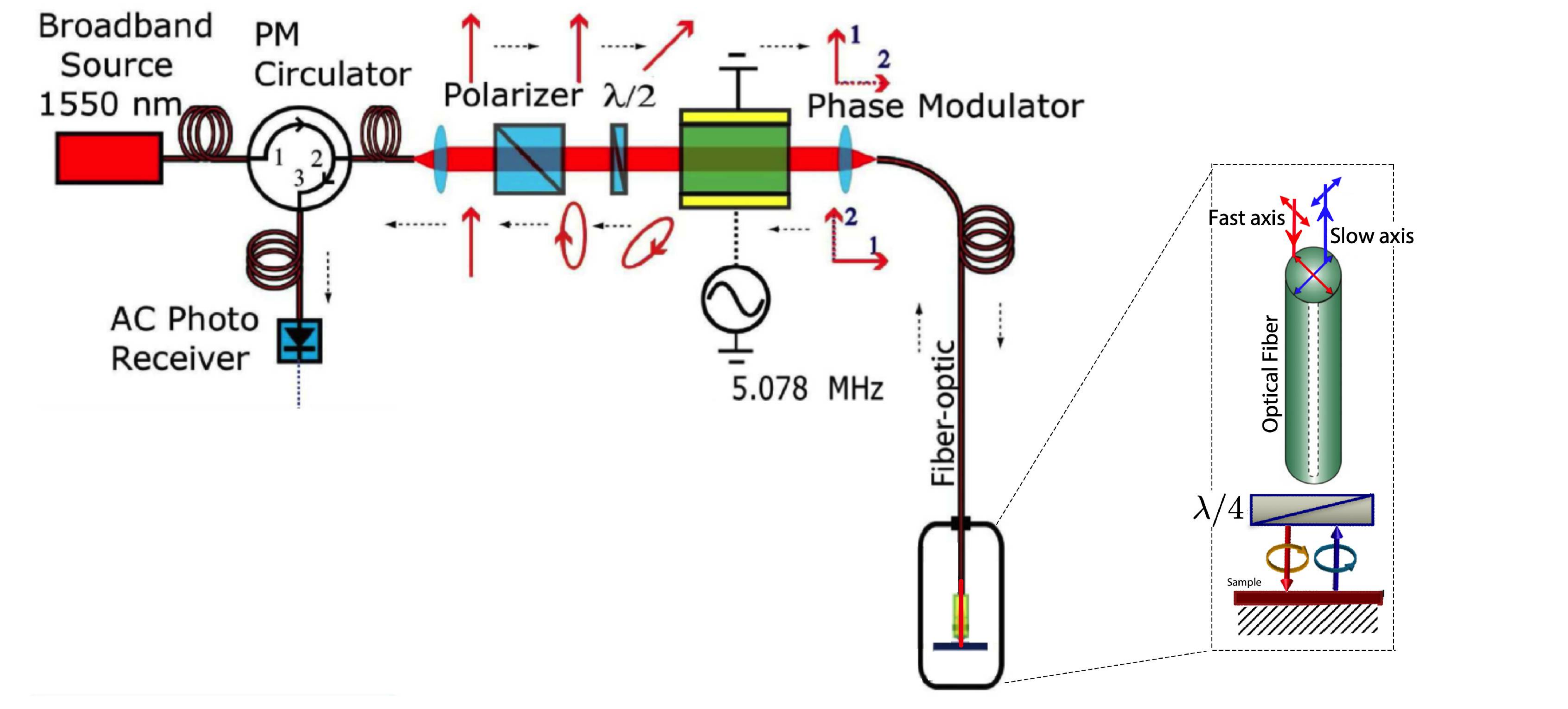}
	\caption{ Schematic of the zero-area loop Sagnac interferometer used in this study. On the right is the end-part showing one beam that enters the quarter-waveplate from the fast axis as right circularly polarized, and reflected back into the slow axis as left circularly polarized after experiencing a Kerr angle shift.  A second beam propagates in the direction of the sample in the slow axis and is reflected into the fast axis.}
	\label{zasi}
\end{figure}

High-resolution measurements of Kerr rotation were performed using a fiber-based zero-area loop Sagnac interferometer as first described in Ref.~\cite{Xia2006}. Following the main components shown in Fig.~\ref{zasi},   two linearly polarized counterpropagating beams comprising the interferometer are isolated along the fast and slow axes of $\sim$10 m of polarization-maintaining fiber and are modulated in phase with a frequency that matches the transit time of light through the interferometer. The linearly polarized light traveling along each of these paths passes through a quarter-waveplate and is converted into circularly polarized light just above the sample. In the absence of magnetic field, the apparatus is completely reciprocal by symmetry except for the sample. Thus, upon reflection, one branch of the interferometer acquires a phase shift of $+\theta_K$, while its orthogonally polarized counterpart acquires an opposite phase shift of $-\theta_K$. The two phase shifts are added at the detector and extracted from the ratio of the first and second harmonics of the modulated signal. 
\bigskip

\subsection*{Field-Cool Measurements}

\begin{figure}[h]
\includegraphics[width=1.0\columnwidth]{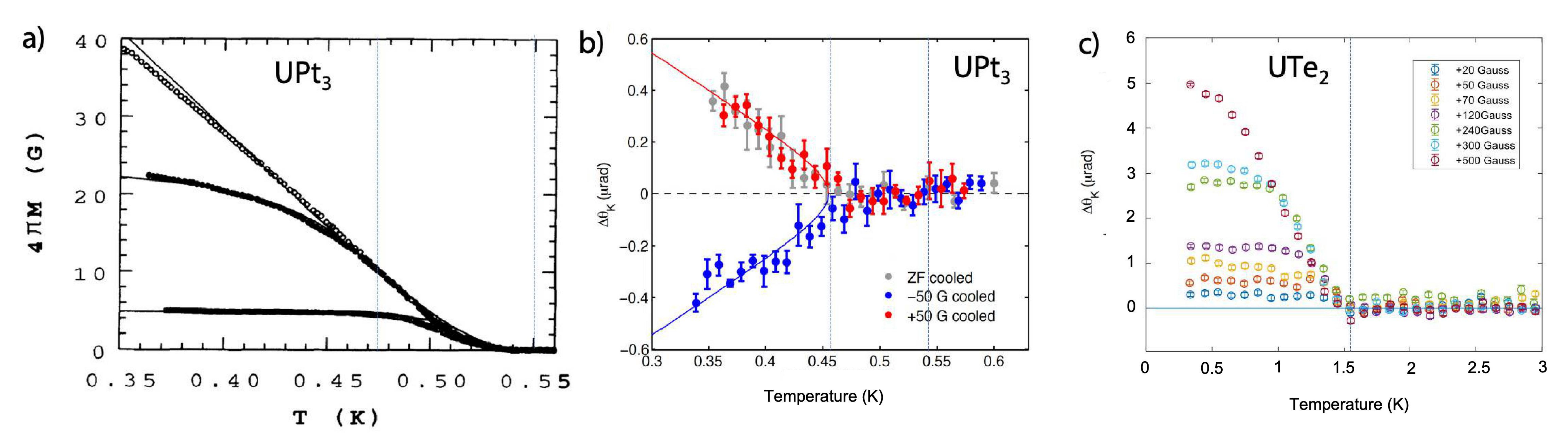}
	\caption{ Field-cool Remanent Magnetization (a) after cooling in magnetic fields of 182, 25 and 5 G (from Ref.~ \cite{Koziol1994}) vs. field-cool Kerr measurements (b) (from Ref.~ \cite{Schemm2014,Schemm_Thesis}) for UPt$_3$ crystals. The $-50$ G FC data was flipped from a negative to positive $\theta_K$. Here solid line is a fit to $\theta_K\propto \sqrt{[1-(T/T_{c-})^2][1-(T/T_{c+})^2]}$, where $T_{c+}=0.55$ K and $T_{c+}=0.47$ K.  Vertical dashed lines mark the two superconducting transitions, with the lower one associated with the TRSB transition as observed by Kerr effect.}
	\label{upt3}
\end{figure}

To compare the field cool observed in UTe$_2$ to other uranium based, but non-paramagnetic systems, we show in Fig.~\ref{upt3} remanent magnetization and Kerr effect data for UPt$_3$. Focusing first on the UPt$_3$ results, we observe that the remanent magnetization in Fig.~\ref{upt3}(a) ``ignores'' the second transition where TRSB is onset for this material, and is fully described a a critical state model with relatively strong pinning \cite{Koziol1994}. By contrast, the Kerr effect in Fig.~\ref{upt3}(b) is independent of whether it was obtained in a field-cool or zero-field-cool measurement, and it starts at the known lower $T_c$ of the material \cite{Schemm2014}. In (c) we show a set of remanent Kerr effect as a function of magnetic field for UTe$_2$, which clearly resemble the behavior of the magnetization of UPt$_3$, but not the Kerr effect response.
\bigskip

\subsection*{Kerr Effect Measurements at Zero-Field}
Kerr angle data was acquired at a rate of 1 sample per second. Due to the slow time constant of the RF lock-in amplifier used to measure the signal, each data point is correlated with its 8 closest neighbors. In order to accurately compute error bars, we eliminated these correlations by dividing the data into 50-point chunks and averaging the points in each chunk, leaving us with a set of averages $\{x_i\}$. These averages are then almost completely uncorrelated and thus represent independent samples with 50 sec. of averaging time. Before plotting the data vs temperature, the chunk averages are then binned into temperature bins, with all the xi in each bin averaged together. The points plotted in all $\theta_K(T)$ graphs are these bin averages. The error bars given are the 1-$\sigma$ standard error of the mean, computed as the standard deviation of the $x_i$ in each bin divided by the square root of the number of chunks.
\bigskip

\subsection*{Kerr Effect Measurements at finite magnetic field}
UTe$_2$ is paramagnetic with a finite positive susceptibility, which works against the Meissner currents in the superconducting state, and against the diamagnetism of the optics (here we note that there are several components of the optics such as a few inches of fiber strand, lenses and quarter waveplate, all transparent insulators and thus exhibit diamagnetism in magnetic field.)  However, since at these low fields both contributions are linear in magnetic field, we can write for the Kerr response in the presence of small magnetic field:
\begin{equation}
\Delta \theta_K = -aH + \chi_K(T) H
\label{kerr}
\end{equation}
where $a$ is the optics diamagnetic contribution and $\chi_K(T)$ is the temperature-dependent Kerr effect due to the magnetic susceptibility of the material in the normal state. 
\begin{figure}[ht]
\subfigure{\label{120G}\includegraphics[width=0.4\columnwidth]{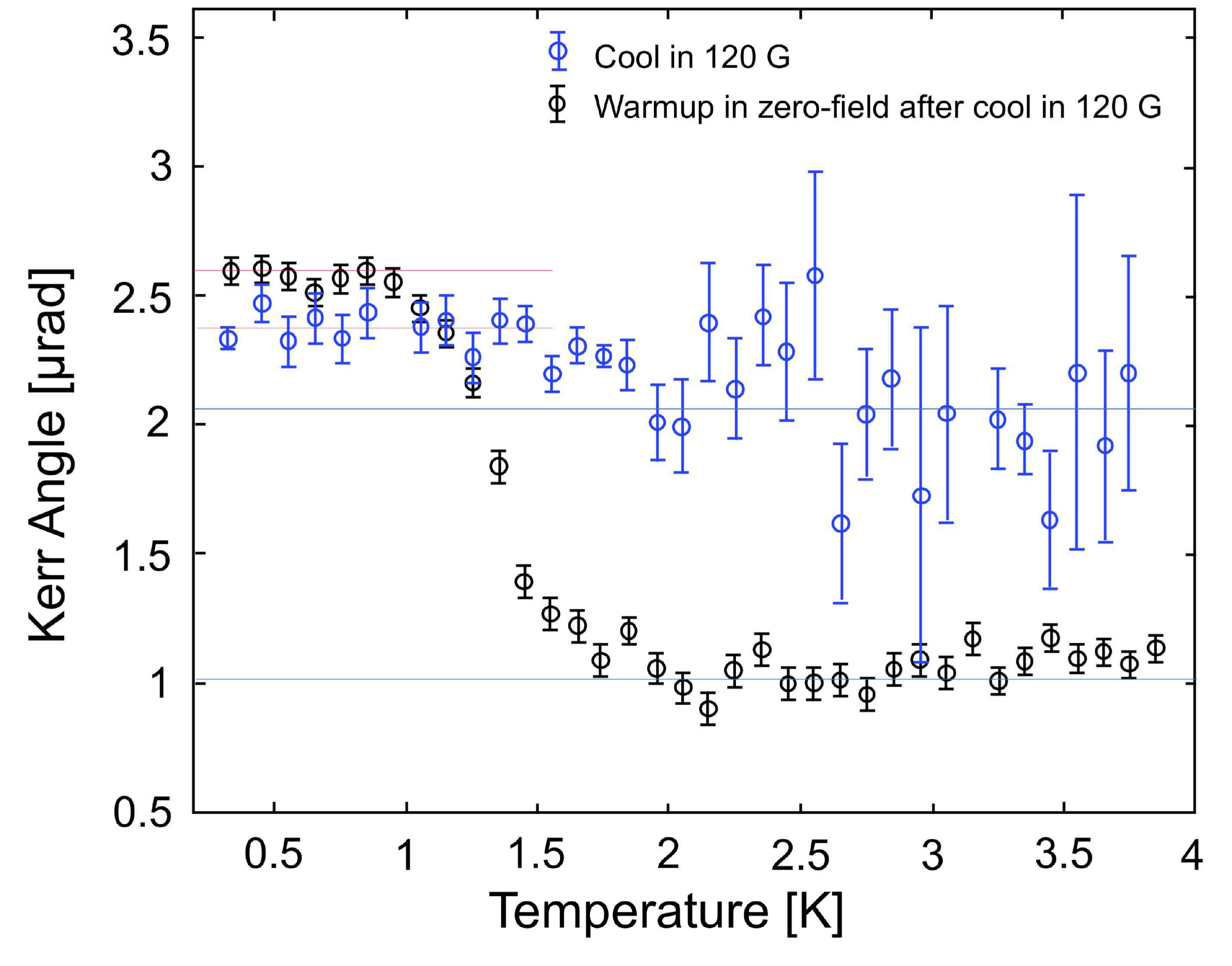}}
\subfigure{\label{240G}\includegraphics[width=0.4\columnwidth]{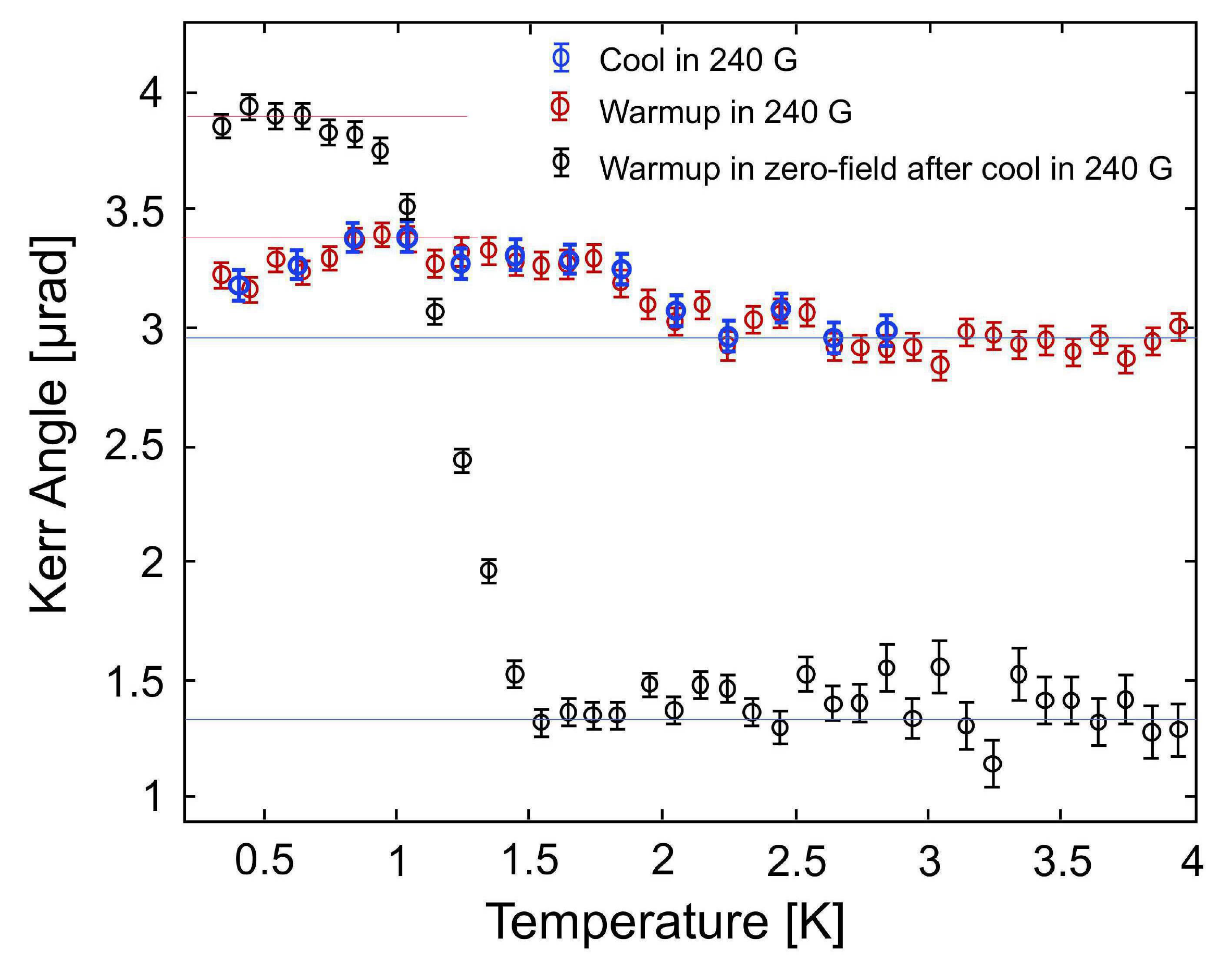}}\\
		\caption{\textbf{(a):} Susceptibility of UTe$_2$ in magnetic fields below $H^*$. \textbf{(b):} Kerr effect measurements. The sample was cooled in magnetic field of 240G and then measured when warmed up in 240G (A)  and in zero-field (B). The vertical red line is the zero-field warmup Tc. }
	\label{infieldX}
\end{figure}
Following our analysis in the previous section, we determined that at e.g. 240G the remanent magnetization is as large as the field applied, and therefore the diamagnetic contribution due to the optics can be extracted as the difference between the in-field and zero-field Kerr angle at base temperature, thus $aH\approx 0.5 \mu$rad as is evidence from Fig.~\ref{infieldX}.

\begin{figure}
\includegraphics[width=0.6\columnwidth]{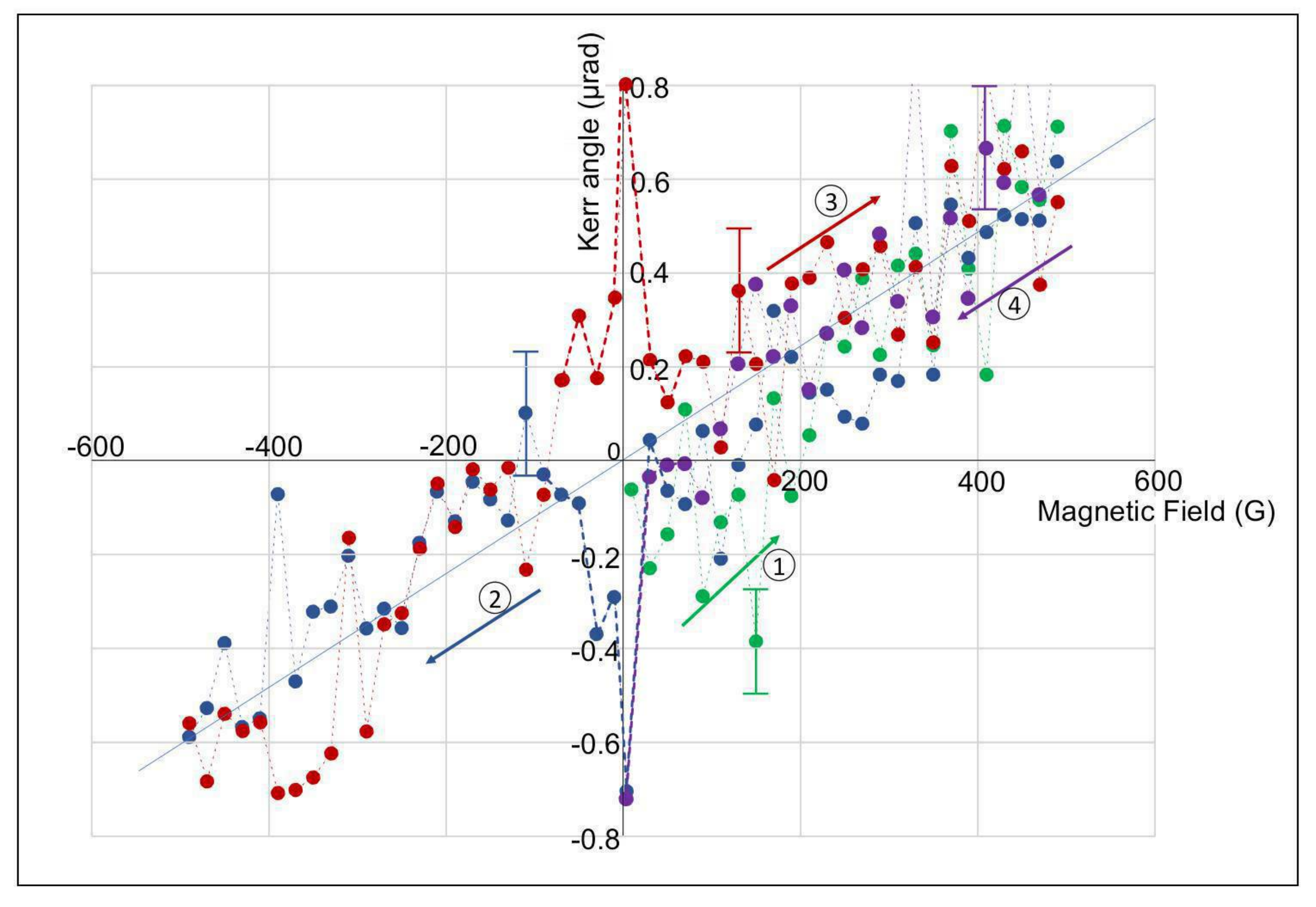}
	\caption{ Hysteresis of Kerr angle between -500 and +500 G. Arrows mark the direction of the field. The multi-looped experiment starts with a ZFC to a value consistent with Fig.~(1)c.  Then in circled section (1) (green markers) the field increased to +500 G, decreased to -500 G in section (2)  (blue markers) , increased back to +500 G in section (3) (red markers) and then back to zero-field in section (4) (purple markers). Note: The optics contribution described in Eqn.~\ref{kerr} was subtracted from the data. The dashed line connect the mean of the data points, and is bolder near zero field to emphasize the increase in Kerr effect.}
	\label{hyst}
\end{figure}

Establishing the effect of the optics, we can subtract this effect from any finite field measurement. In particular we show in Fig.~\ref{hyst} a hysteresis loop performed at 0.3 K in the magnetic field range of $\pm 500$ G. The sample was first cooled to base temperature in zero magnetic field. As explained in the main text, in such ZFC the observed Kerr effect will typically lie in the range of $\pm0.4~\mu$rad, where smaller Kerr angles correspond to domain averaging effect. The Kerr effect was then measured while increasing the magnetic field to $+500$ G, then decreasing it to $-500$ G, increasing it to $+500$ G, and finally decreasing it to zero. While the scatter in the data is rather large, presumably due to domain effects and vortex motion, it is clear that unlike the field-cool data shown above, after cooling in zero field the Kerr response is weak, indicating the good shielding of the sample in preventing vortices to reach the center of the sample (note: the Kerr effect is measured with a beam of size $\sim 10~\mu$m at the center of the $\sim 1.2$ mm size sample.) At the same time we clearly observe the sharp increase in Kerr angle as we approach zero magnetic field, indicating the critical current associated with the presence of trapped flux, which needs to change direction as the field is reversed. Compared to the field cool measurements at 500 G (Fig.~4 in the main text), we see that only 15\% of the flux is trapped at the center of the sample where Kerr effect is measured. Below $\sim 250$ G similar loops yield a zero-field intersect of $<0.4~\mu$rad, indicating a more intrinsic effect, which we argue to be the effect of the order-parameter itself. 
\bigskip

\subsection*{Magnetic Susceptibility}
Figure~\ref{susceptibility} below shows the low-temperature part of the magnetic susceptibility of a UTe$_2$ crystal measured in a SQUID magnetometer down to 1.8 K.  Focusing on the $c$-axis, Fig.~\ref{susc1} indicates that the low field susceptibility is very weakly field dependent, with a shallow minimum at $\sim240$ G. This plot is therefore important in comparing the susceptibility to the Kerr effect that we analyze in the main text (See Figs.~5 and 7.) Fig.~\ref{susc2} is the magnetic susceptibility along the three principal axes of the crystal, showing a weaker than Curie-law for either direction. 
\begin{figure}[h]
	\centering
	\subfigure{\label{susc1}\includegraphics[width=0.35\columnwidth]{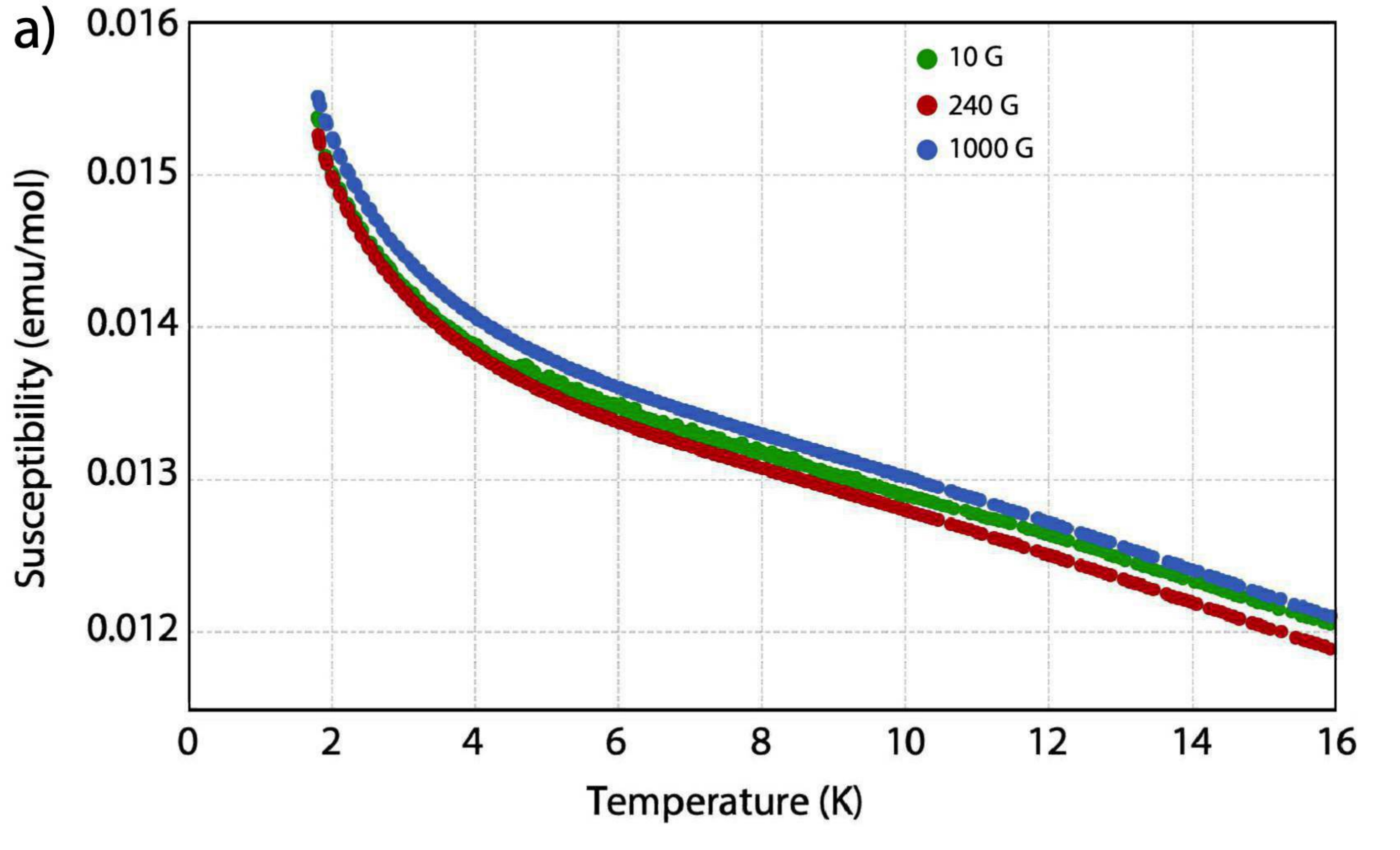}}
\subfigure{\label{susc2}\includegraphics[width=0.33\columnwidth]{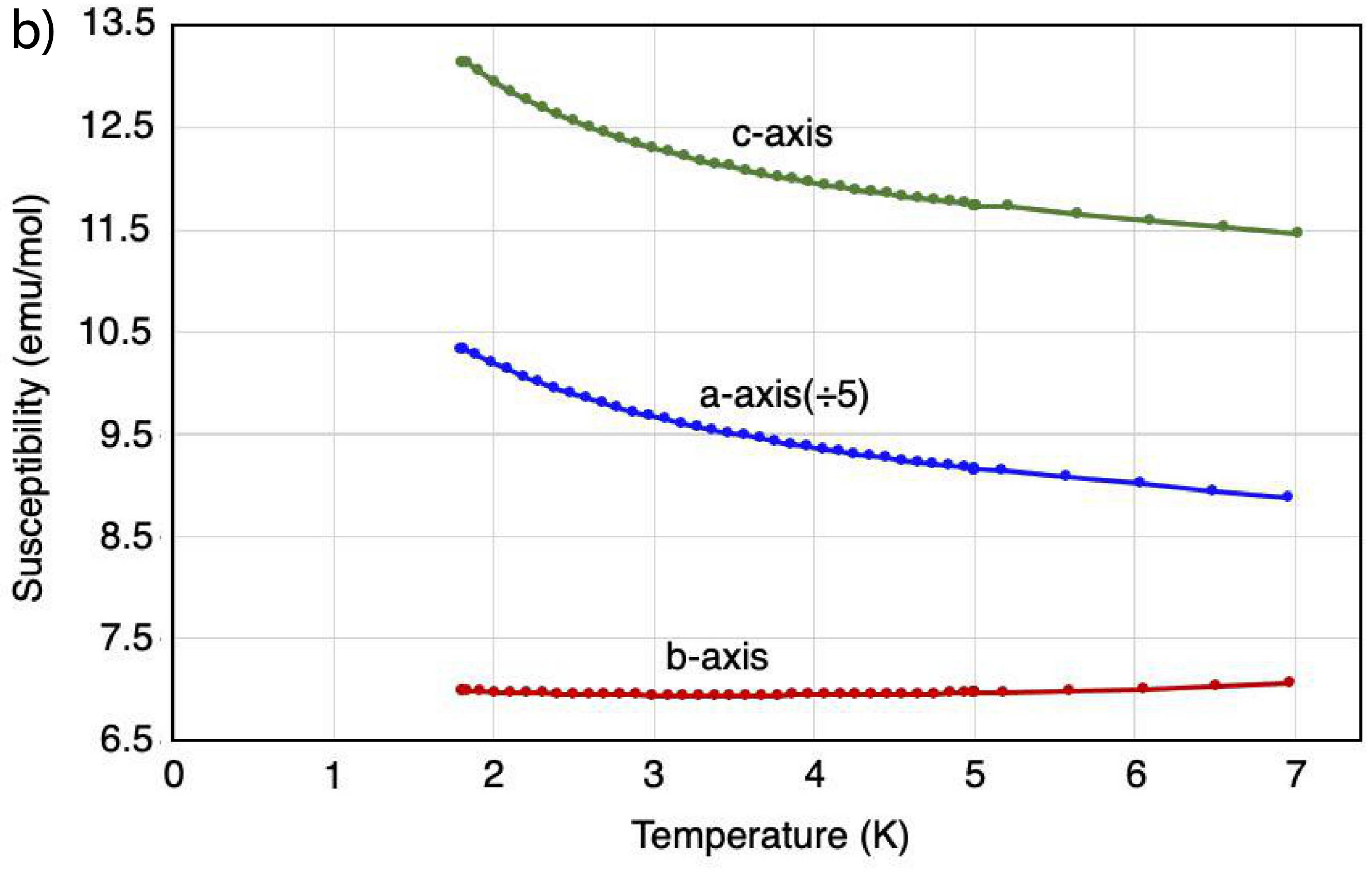}}\\
	\caption{Magnetic susceptibillity of UTe$_2$ single crystal above $T_c$. (a) Measurements along the $c$-axis at 10 G, 240 G and 1000 G, showing weak magnetic field dependence and a shallow minimum in the field range of $\sim$240 G. (b) Susceptibility at 1000 G. Note that $a$-axis (easy axis) response is largest, while along the $b$-axis a minimum around 3 K is barely visible. }
		\label{susceptibility}
\end{figure}
\bigskip

\subsection*{The Lower Critical Field of UTe$_2$}

A first task is therefore to determine the lower critical field  $H_{c1}$, which for UTe$_2$ is not simple. Low field magnetization isotherms were reported by Paulsen {\it et al.} \cite{Paulsen2021} along the $a$- and $b$- axes, indicating a reasonable fit to the ``standard'' form $H_{c1}(T)=H_{c1}(0)[1-(T/T_c)^2]$, with $H^a_{c1}(0)=23$G and $H^b_{c1}(0)=42$G, and the good fits could be an indication of the reliability of the measurement.  However, these measurements do not seem to agree with complementary measurements of the thermodynamic and upper critical fields, expecting:
\be
H_{c1}H_{c2}=H_c^2[{\rm ln}\kappa +c],
\ee
where $\kappa$ is the Ginzburg-Landau (GL) parameter, and $c\approx 0.5$ is a constant. In fact, a test of the $H_{c1}(0)$ values against the separately measured upper-critical fields $H_{c2}(0)$, and thermodynamic critical field $H_c(0)$ indicate that only the $a$-axis direction agrees with the expected relation. Furthermore, the anisotropy of $H_{c1}$ along $a$- and $b$- axes is found to be opposite to that expected from the anisotropy of the respective $H_{c2}(T)$, even close to $T_c$ \cite{Paulsen2021}. These results might indicate that estimating $H^c_{c1}(T)$ from standard data may be complicated. For example, using the thermodynamic critical field, $H_c(0)= 490$G \cite{Paulsen2021},  and $H^c_{c2}(0)= 7.5$ T (as extrapolated from low-field in \cite{Ran2019}), and a  GL-parameter of $\kappa \approx 110$, estimated separately from $H_{c2}(0)=\sqrt{2}\kappa H_c(0)$, we obtain  an estimate of $H^c_{c1}(0)\approx 17$G, while in the same paper \cite{Paulsen2021}, using remanent magnetization arguments including demagnetization factors, a value of $\sim 13$ G was deduced.  At the same time, Bae {\it et al.} \cite{Bae2021} estimated the effective penetration depth $\lambda^{ab}_{eff}$, when a magnetic field is applied along the $c$-axis to be in the range of 1126 nm to 947 nm. Taking the shorter value gives a maximum estimate of $H^c_{c1}(0)\approx 8.5$G. Turning to our own data of the value of the low-temperature saturated Kerr effect taken from Figs.~\ref{lowfield} and~\ref{fields} and plotted in Fig.~\ref{slope}, we also find that $10 {\rm G}\lesssim H^c_{c1}(0) \lesssim 25 {\rm G}$. All these results suggest that it is safe to assume $H_{c1}(0)\lesssim 20$G, which will be important for the discussion below.\\

\end{document}